\documentclass[12pt]{iopart}

\usepackage{graphicx}
\usepackage{amssymb}
\usepackage{hyperref}

\begin{document}

\title[A Rubidium Vapor Source ...]{A Rubidium Vapor Source for a Plasma Source for AWAKE}

\author{G. Plyushchev$^{1,2}$, R. Kersevan$^{1}$, A. Petrenko$^{1}$, P. Muggli$^{3,1}$}

\address{$^1$CERN, Geneva, Switzerland\\$^2$Ecole Polytechnique F{\'e}d{\'e}rale de Lausanne, Switzerland\\$^3$Max Planck Institute for Physics, M{\"u}nchen, Germany}
\ead{muggli@mpp.mpg.de}
\vspace{10pt}
\begin{indented}
\item[]July 23, 2017
\end{indented}

\begin{abstract}
We present the scheme for a rubidium vapor source that is used as a plasma source in the AWAKE plasma wakefield acceleration experiment. %
The plasma wakefield acceleration process requires a number of stringent parameters for the plasma: electron density adjustable in the (1-10)$\times$10$^{14}$\,cm$^{-3}$ range, 0.25\% relative density uniformity, sharp ($<$10\,cm) density ramps at each end, density gradient adjustable from -3 to +10\% over 10\,m, and \%-level density step near the beginning the plasma column. %
We show with analytical and direct Simulation Monte Carlo results that the rubidium density in the proposed source should meet these requirements. %
Laser ionization then transfers the above neutral vapor parameters to the plasma. %
\end{abstract}

%
%
\submitto{\PSST}
%
%
%
\section{Introduction}

Relativistic charged particle bunches can drive large accelerating fields in plasmas known as plasma wakefields~\cite{bib:chen}. %
In order to reach field amplitudes larger than 1\,GV/m the plasma electron density must exceed 10$^{14}$\,cm$^{-3}$. %
To effectively drive the wakefields, the bunch must be smaller than the period of the wakefields in all its dimensions, that is smaller than $\sim$3.3\,mm in this case. %

Proton bunches produced for example by the CERN super proton synchrotron (SPS) or large hadron collider (LHC) carry much more energy, 10s to 100s of kilojoules, than electron bunches usually used to drive wakefields (10 to 100 joules). %
They can thus in principle drive wakefields over a much longer plasma length than electron bunches can. %
Therefore, they could be used to accelerate a witness electron bunch to very high energies in a single, long plasma section, e.g., 500\,GeV over $\sim$500\,m~\cite{bib:caldwellshort}. %

Proton bunches can be focused to a small transverse size, $\sim$200\,$\mu$m, but they are long, 10-12\,cm. %
The wakefields ineffectively driven by the long bunch can self-modulate it into a train of bunches with length shorter than, and period equal to that of the wakefields~\cite{bib:kumar}. %
The train of many short bunches naturally formed can then resonantly drive the wakefields to large amplitudes. %
A witness electron bunch can be injected into the accelerating and focusing phase of the wakefields and reach high energies~\cite. %

The self-modulation process, the resonant excitation of the wakefields and the electron acceleration process put stringent requirements on the plasma characteristics and thus on the vapor source. %
The source that satisfies these requirements consists of a 10\,m-long rubidium (Rb) vapor column ionized by a $\sim$100\,fs-long, terawatt power level laser pulse~\cite{bib:ozdensity}. %
This source is one of the key elements of the AWAKE experiment~\cite{bib:AWAKE} at CERN. %

In this paper, we explain the requirements for the plasma and vapor source. %
We describe the Rb vapor source and show analytical and direct simulation Monte Carlo (DSMC) results indicating that the source satisfies the necessary requirements. %
The ionization, that turns the vapor characteristics into electron plasma density characteristics, will be described in another publication. %
Hereafter, gas, vapor and plasma densities are often used interchangeably. %

\section{Vapor Source Requirements}
To reach GV/m accelerating field amplitudes the plasma electron density must be adjustable in the (1-10)$\times$10$^{14}$\,cm$^{-3}$. %
Assuming for now that the laser pulse ionizes only the first electron of all atoms in its path, the atomic density must be in the same range. %

Numerical simulation results~\cite{Caldwell16} show that the self-modulation process takes around 4\,m to saturate. %
Therefore the plasma length must be $\sim$10\,m to accommodate for some acceleration of externally injected electrons. %
At the above plasma electron densities, the transverse size of the wakefields is expected to be $<$1\,mm. %
However, to give clearance for the high-energy proton bunch, the vacuum pipe diameter must be $\sim$4\,cm. %

The proton micro-bunches, resulting from the self-modulation process, resonantly excite the wakefields. %
For the process to be effective, the plasma density variations along the 10\,m plasma must not exceed $\cong$1/N$\cong$1\% for N$\cong$100 short bunches, typical for the $\sim$12\,cm-long proton bunch and 1.2\,mm wakefields period. %
In addition, the electrons to be accelerated must reside in the phase of the wakefields where they are both accelerated and focused. %
In linear theory, this is over one quarter of the wakefields period, bringing the maximum density variation requirement to $\frac{\delta n}{n}\cong\frac{1}{4}\frac{1}{N}$=0.25\%. %

The low energy ($\cong$10-20\,MeV) electrons must enter the plasma through the density ramp located at its entrance. %
In this region, the wakefields are globally defocusing for the electrons and their period varies rapidly, with the square root of the local density, and their amplitude also varies. %
The length of the ramp must therefore be kept as short as possible, $<$10\,cm according to numerical simulations~\cite{Lotov13}. %
Since electrons exit the source at much higher energies ($\cong$1\,GeV) than they enter it, they are much less sensitive to the effect of the transverse wakefields in the exit density ramp. %

The three beams (proton, laser, electron) with very different characteristics must be able to enter and exit the source without loss, scattering or damage caused. %

Optimal electron acceleration may require a density gradient along the source, between -3\% to +10\%~\cite{Caldwell16}. %
We note here that optimum driving of the wakefields and optimum acceleration of electrons may lead to seemingly mutually exclusive requirements: uniformity and gradient. %
This is due to the fact that the final electron energy is proportional to the integral of the accelerating field experienced by the particles along the plasma. %
During acceleration electrons and wakefields may dephase. %
Some of this dephasing can be compensated for by changing the wakefields period along the source, through the density gradient. %
A plasma with a density gradients (over a finite distance) can therefore lead to larger energy gain than one of constant density. %

The development of the self-modulation process requires a small density step at the \%-level over some centimeters in the first meters of the plasma~\cite{Caldwell16} for the wakefield amplitude to be maintained over long distances. %

\section{Vapor Source}

As mentioned before, the source that satisfies these requirements consists of a 10\,m-long Rb vapor column ionized by a laser pulse~\cite{bib:ozdensity}. %

Rubidium was chosen because of the low ionization potential of its first electron (4.177\,eV), making it relatively easy to field-ionize with an intense laser pulse of moderate focused intensity, $\sim$10$^{12}$\,Wcm$^{-2}$. %
Neutral densities in the range of (1-10)$\times$10$^{14}$ Rb atoms per cubic centimeter can be obtained by evaporating the alkali metal at temperatures between about 150 and 200$^o$C, thanks to the high vapor pressure of Rb. %
Rubidium has a melting temperature of 39.48$^o$C, which means that it condenses on vacuum chamber walls maintained at room temperature. %
Rubidium atoms also have a relatively large atomic mass. %
Natural Rb comprises two isotopes: 72\% is the stable isotope $^{85}$Rb, and 28\% is $^{87}$Rb the unstable isotope with a 49 billions years half-life. %
The relatively heavy ion mass mitigates possible plasma ions motion that could interfere with the acceleration process and with the maintaining of the electron bunch emittance~\cite{bib:vieiraion}. %

\begin{figure}[h!]
\centering
\includegraphics[width=0.7\textwidth]{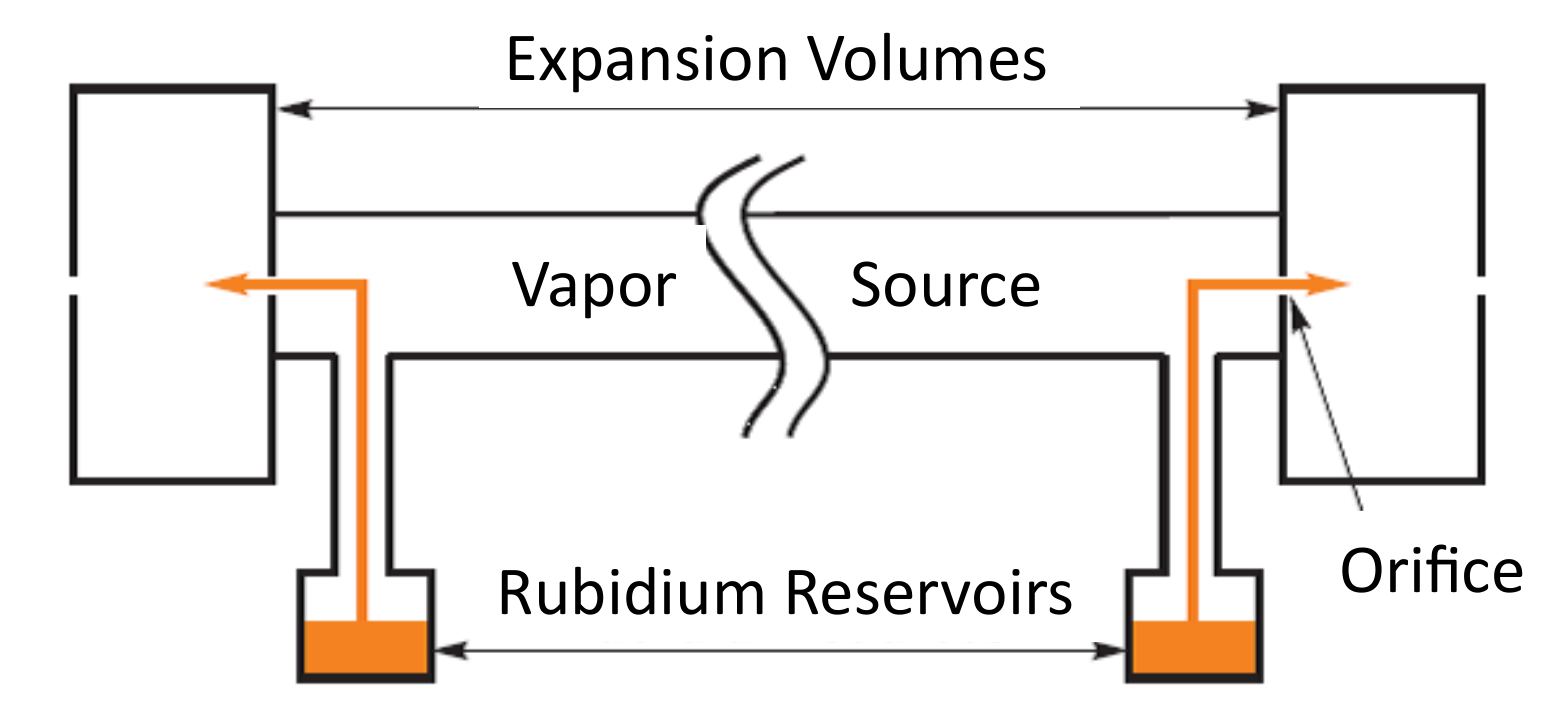}
\caption{Schematic of the AWAKE vapor source. The orange arrows indicate the Rb vapor flow from the Rb reservoirs to the expansion volumes. (Not to scale!)}
\label{fig:design}
\end{figure}

Figure~\ref{fig:design} shows a schematic of the Rb vapor source. %
In absence of significant flow, imposing a very uniform temperature along the source tube leads to a very uniform vapor density: $\delta n/n=\delta T/T$. %
Reservoirs located at both ends evaporate the Rb that fills the source tube. %
The source tube has a thin (600\,$\mu$m), on-axis, 10\,mm diameter circular orifice at each end, where the Rb vapor escapes and expands into a much larger expansion volume kept under vacuum, $\sim$10$^{-7}$\,mbar. %
Expansion and condensation of the Rb on the nearby walls kept at room temperature provide the sharp density ramp. %
The orifices are of course transparent for the three beams. %
We use the adjustment of the reservoirs temperature to set the density gradient along the source (including no gradient). %

\section{Rubidium Flow Characteristics}

We describe here the type of flow that can be expected for the Rb vapor with the parameters of interest here. %

The mean free path of the Rb atoms $\lambda_{mfp}$ in a vapor of density n, temperature T and pressure $p=nk_BT$ can be calculated from the diameter of the atom d and from the vapor viscosity $\mu$: 
\begin{equation}
\lambda_{mfp}=\frac {k_B T}{\sqrt{2}\pi d^2 p}=\frac{\mu}{p}\sqrt{\frac {\pi k_B T}{2 m}}.
\end{equation}
Using these equations we can calculate the dynamic viscosity as:
\begin{equation}
\mu=\frac {1}{\pi^{3/2}} \sqrt{m k_B T} \frac {1}{d^2}.
\end{equation}
It should be noted that this equation gives approximately the same result as the hard sphere approximation for viscosity:
\begin{equation}
\mu=\frac {5}{16} \sqrt{\frac {m k_B T}{\pi}} \frac {1}{d^2}
\end{equation}
At a temperature of 500\,K, with the Rb atom mass m=1.419$\times$10$^{-25}$\,kg and diameter d=496\,pm, the viscosity is $\mu$=2.3$\times$10$^{-5}$\,Pa$\cdot$s. %
The mean free path for a density of $7\times$10$^{14}$\,cm$^{-3}$ is thus $\lambda_{mfp}$=1.31\,mm.

The rarefaction parameter is another quantity that characterizes the flow. %
It is inversely proportional to the Knudsen number and is defined as:
\begin{equation}
\delta=\frac {a}{l}
\label{eq:rarfac}
\end{equation}
where l is the equivalent molecular mean free path ($l=\frac {2}{\sqrt{\pi}}\lambda_{mfp}$). %
Also:
\begin{equation}
l=\frac {\mu v_m}{p}=\frac{\mu}{p}\sqrt{\frac{2 k_BT}{m}}
\end{equation}
where $v_m=\sqrt{\frac{2 k_B T}{m}}$ is the representative most probable velocity of the atoms. %
The Knudsen number Kn is defined as:
\begin{equation}
Kn=\frac{\lambda_{mfp}}{a}
\end{equation}
where a is the radius of the circular pipe containing the vapor. %

There are three distinct flow regimes, depending on the importance of collisions between the atoms. %
The continuum regime is characterized by a high collision rate and occurs at high density. %
The mean free path in this case is much smaller than the characteristics size of system. %
In the other limiting case, when the density is very low, the collisions between atoms are negligible as the mean free path is much longer than the characteristics size of the system. %
This is the molecular regime. %
The intermediate regime is called the transitional regime. %
This regime is characterized by a Knudsen number between 0.1 and 10. %
For the case of a=2\,cm, Kn$\cong0.07$ and the flow is between the molecular and the transitional regime for the range of densities of interest here. %

The DSMC method~\cite{bib:dsmc} is therefore appropriate to describe the flow situation of the vapor source. %
Initially the Test-Particle Montecarlo code (TPMC) Molflow+~\cite{kersevan} has been used to simulate the process of Rb density gradient formation~\cite{kersevan2}. %
Molflow+ is capable of reproducing the analytical density profile as in Fig.~\ref{fig:den1ddsmc}. %
At a later stage the more appropriate DSMC method has been employed. %

In the following we present DSMC results and analytical results, when possible. %

\section{Density Ramp}

In this Section we discuss the characteristics of the vapor in the density ramps at the ends of the vapor column. %

\subsection{Density Ramp Length}
\begin{figure}[h!]
\centering
\includegraphics[width=0.8\textwidth]{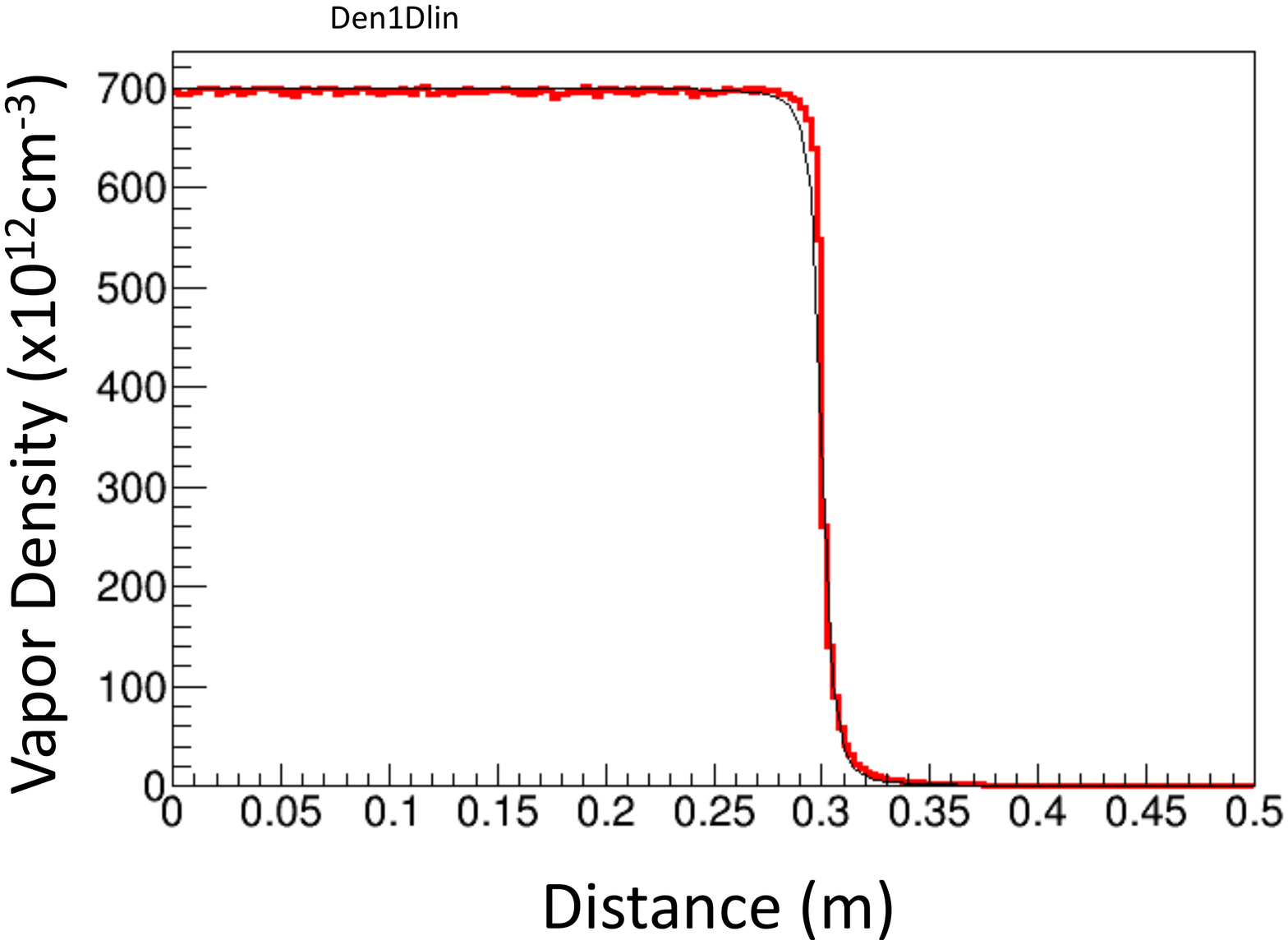}
\caption{Thin black line: plot of Eq.~\ref{eq:densorif}. %
Thick red line: DSMC result. %
Complete condensation is assumed on all the walls, except for the orifice wall in the expansion volume where the vapor expands. %
Orifice located 0.3\,m, vapor column on the left hand side, expansion volume on the right hand side of the orifice.}
\label{fig:den1ddsmc}
\end{figure}

In the molecular regime, the density distribution on the line perpendicular to the orifice plane and at the centre of orifice opening, i.e., along the source axis is~\cite{danilatos}:
\begin{equation}
\frac {n(x)}{n_0}=\frac {1}{2}-\frac {(x/2r)}{ \sqrt{(x/2r)^2+0.25}}
\label{eq:densorif}
\end{equation}
Here r is the orifice radius and n$_0$ is the density far inside the tube. %
Thus the gradient length is on the order of the orifice radius, as can be seen from the plot of Eq.~\ref{eq:densorif} on Fig.~\ref{fig:den1ddsmc}. %
This density distribution is perfectly symmetric with respect to the orifice plane (x=0 in Eq.~\ref{eq:densorif}, 0.3\,m on Fig.~\ref{fig:den1ddsmc}). %

Direct simulation Monte Carlo were performed in order to confirm the scale of the ramp length. 
In the simulations we use a hard-sphere collisions model with a Rb atom diameter of 496\,pm. %
This approximation is reasonable in the case of the monoatomic gas. %
The on axis density form the simulations is also shown on Fig.~\ref{fig:den1ddsmc}. %
The ramp length is indeed only a few centimeters and in good general agreement with the analytical expression of Eq.~\ref{eq:densorif}. %

In reference~\cite{sharipov01} the density distribution near an orifice is presented for the rarefied gas regime and for different rarefaction parameters. %
Here the orifice radius rather than the vapor source radius should be used in the definition of the rarefaction parameter. %
Equation~\ref{eq:densorif} is valid up to a rarefaction parameter $\delta<$1.0. %
In our case, for the 1\,cm orifice diameter and the 7$\times$10$^{14}$\,cm$^{-3}$ density at 500\,K, the rarefaction parameter $\delta$=3.4 and this equation should be used as a good indication for the real ramp length. %
Indeed, on the vapor source side, the simulated ramp is a little bit shorter than the analytical expression predicts (see Fig.~\ref{fig:den1ddsmc}). %
This is due to the fact that with our parameters, the expansion is in the transitional regime and not in the molecular one, as assumed for the analytical formula.

\subsection{Mass Flow through the Orifices}
The expression for the mass flow rate through thin a orifice in different flow regimes is~\cite{sharipov01}:
\begin{equation}
\dot M=Wr^2 n \sqrt{ \frac {\pi mk_BT}{2}}
\label{eq:dotMorf}
\end{equation}
The empirical equation for $W$ is:
\begin{equation}
W=1+\frac{0.4733+0.6005/ \sqrt{\delta}}{1+4.559/\delta+3.094/\delta^2}
\label{eq:worf}
\end{equation} 
and takes into account the rarefaction parameter $\delta$. %
The parameter W normalizes the mass flow rate to the molecular value: for a continuum flow W=1.5, and for a molecular flow W=1.0. %
One can see that in the molecular flow regime the mass flow rate through an orifice is linear with density. %
It is not linear with density in the intermediate flow regime, due to the value of the parameter W.
In our case, W=1.31, which gives $\dot M$=0.90\,mg/s (for $\mu$=2.3 $\times$10$^{-5}$\,Pa$\cdot$ s). %

The simulation result gives $\dot M$=(0.83$\pm$0.01)\,mg/s. %
The small difference with the analytical formula could be explained by the uncertainty on the viscosity value. 

\subsection{Density Perturbation}
\begin{figure}[h!]
\centering
\includegraphics[width=0.9\textwidth]{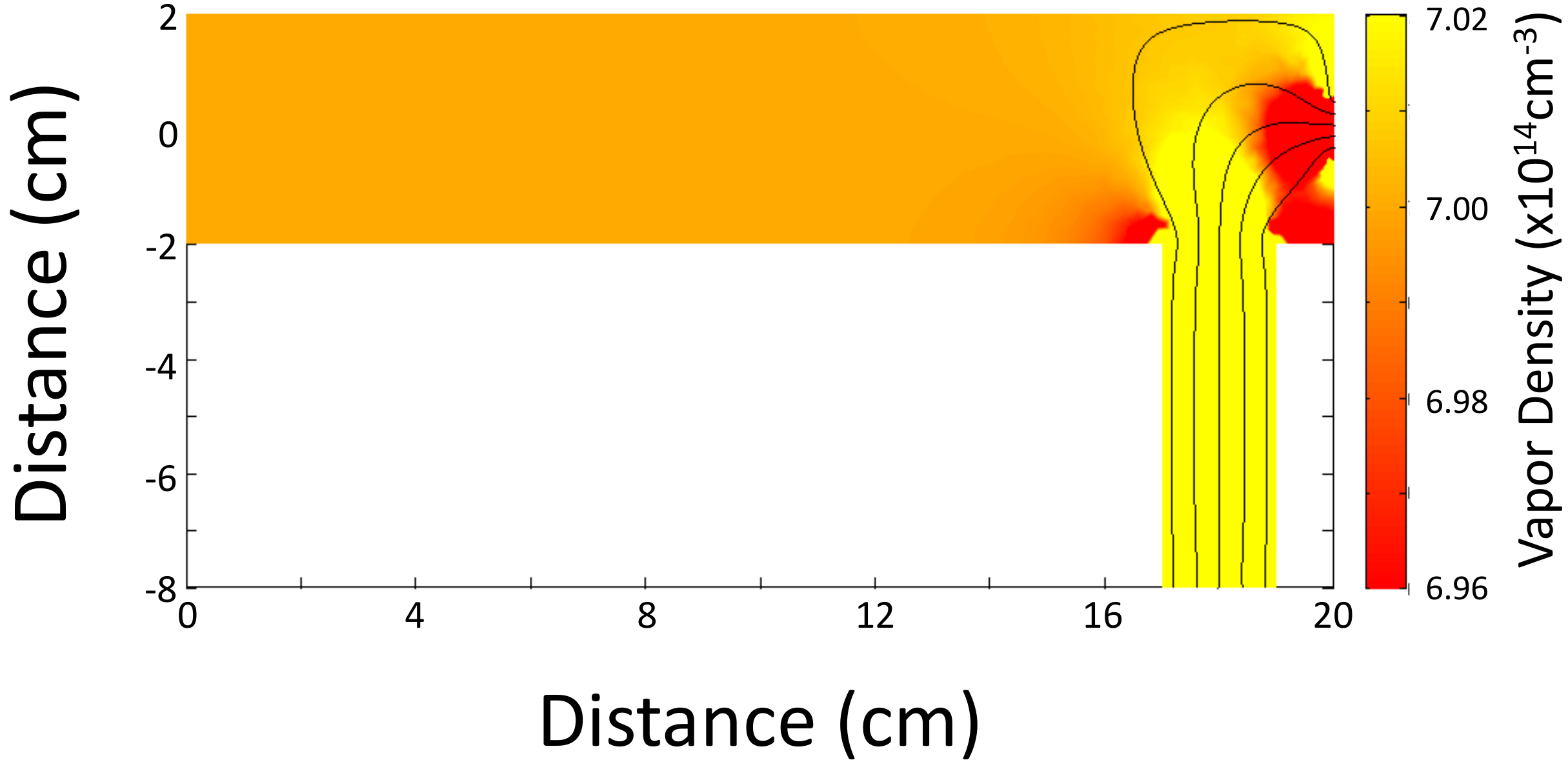}
\caption{COMSOL simulation result for the density profile near the orifice region inside the vapor source. %
Black lines shows the streamlines of Rb flow from Rb reservoir to the orifice opening. %
The color scale is chosen to display small density variations.
Orifice and expansion volume located on the very left of the figure.}
\label{fig:comsol2d}
\end{figure}

Because of the stochastic nature of the simulations, it is very difficult to capture small density variations with DSMC. %
Therefore, we used the Finite Element Method software COMSOL Multiphysics in order to qualitatively investigate the details of the density variations near the orifice inside the vapor source (in front of the injection tube see Fig.~\ref{fig:design}). %
One can expect that the vapor flow from the Rb reservoirs perturbs the density locally. %
This is confirmed by Fig.~\ref{fig:comsol2d} that shows the vapor density distribution near the Rb reservoir tube and source orifice, as obtained with the COMSOL software. %
Note that we use the geometry of the actual vapor source. %
The flow of Rb from the reservoir exits a 6\,cm in length, 1\,cm in radius circular tube, whose axis is 2\,cm away from the source orifices (see Fig.~\ref{fig:comsol2d}). %
The simulations were performed for the continuum flow regime. %
As there is a rarefied flow in this region, the simulations give only qualitative results. %
A line out along the source axis (not shown) shows a small density bump, less than 0.5\%, over distance of $\sim$4\,cm in front of the reservoir tube. %

\section{Vapor Source Operation without Density Gradient}

The results presented here above assume that the source is operated without density gradient and in steady state, i.e., the two ends are identical. %
Therefore, only the few centimeters near the orifices ned to be simulated. %
In this case, the flow is between the Rb reservoirs and the source orifice and into the expansion volume with a mass flow value of $\sim$0.90\,mg/s. %
This flow causes a very slight density perturbation right in front of the reservoir tube and orifice. %
The vapor flow through the orifices results in a density ramp over a length of the order of the orifice diameter, a few centimeters in this case. %
Here after we examine the case of the source operating with a global density gradient as well as the effect of the finite size expansion volume on the density ramp. %
We also examine the Rb reservoir parameters to generate the flows calculated here. %

\section{Operation with Density Gradient, Flows}

As an extreme example, we determine the Rb reservoirs parameters to drive sufficient vapor flows to maintain a +10\% gradient along the source, the maximum expected value for AWAKE. %
A positive gradient means increasing density along beams propagation direction. %

\subsection{Flows along the Source}
\begin{figure}[h!]
\centering
\includegraphics[width=0.8\textwidth]{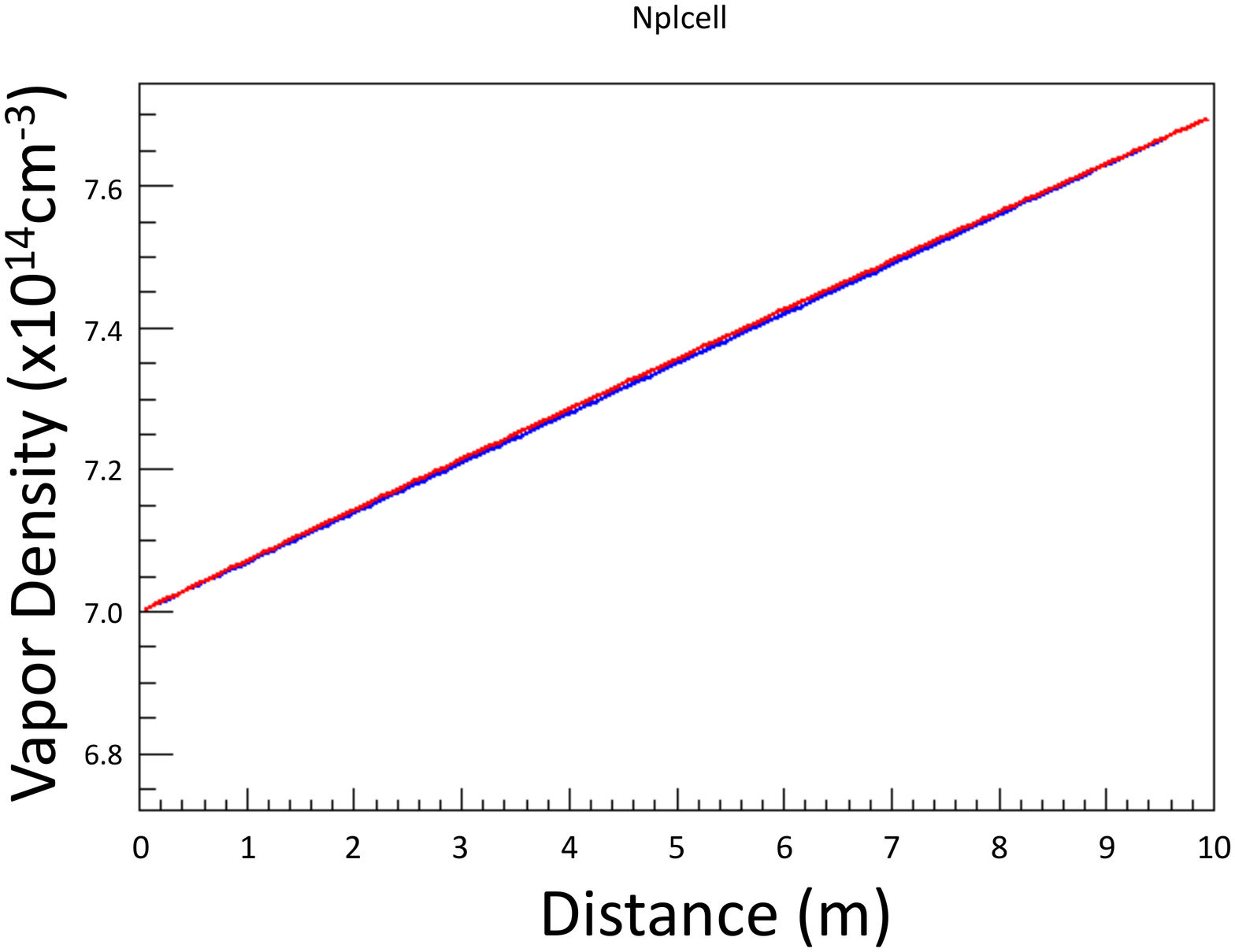}
\caption{Density profile along the vapor source: red line from Eq.~\ref{eq:naccur} and blue line its
linear approximation.}
\label{fig:Nplcell}
\end{figure}

Let us consider a +10\% density gradient along the vapor source with n$_0$=7.0$\times$10$^{14}$ and n$_1$=7.7$\times$10$^{14}$\,cm$^{-3}$ at the source entrance and exit, respectively. %
Using Poiseuille law for the rarefied regime~\cite{sharipov98,sharipovCERN} one finds:
\begin{equation}
n+n_a=\sqrt{ \left( n_0+n_a\right)^2+\frac{x}{L}\left(\left(n_1+n_a\right)^2-\left(n_0+n_a\right)^2 \right)}
\label{eq:naccur}
\end{equation}
\begin{equation}
n_a=\frac {4 \mu v_m \sigma_p}{ak_BT}
\label{eq:nadef}
\end{equation}
where $\sigma_p$ is the viscous slip coefficient. %
This approximation works as long as the tube diameter (4\,cm) is much shorter than the tube length (10\,m). %
For our parameters T=500\,K and $\sigma_p$=1.018: n$_a$=2.102$\times$10$^{14}$\,cm$^{-3}$. %
As the gradient is small, that is $n_1-n_0 \ll2n_0$, the density profile along the vapor source is linear with a good level of approximation, as can be seen on Fig.~\ref{fig:Nplcell} where it is compared to the linear expectation. %
The mass flow rate induced along the source by this +10\% density gradient is $\dot M_{PC}$ =0.0178\,mg/s. 
This is also the Rb mass flow imbalance between the two orifices. %

\begin{figure}[h!]
\centering
\includegraphics[width=0.8\textwidth]{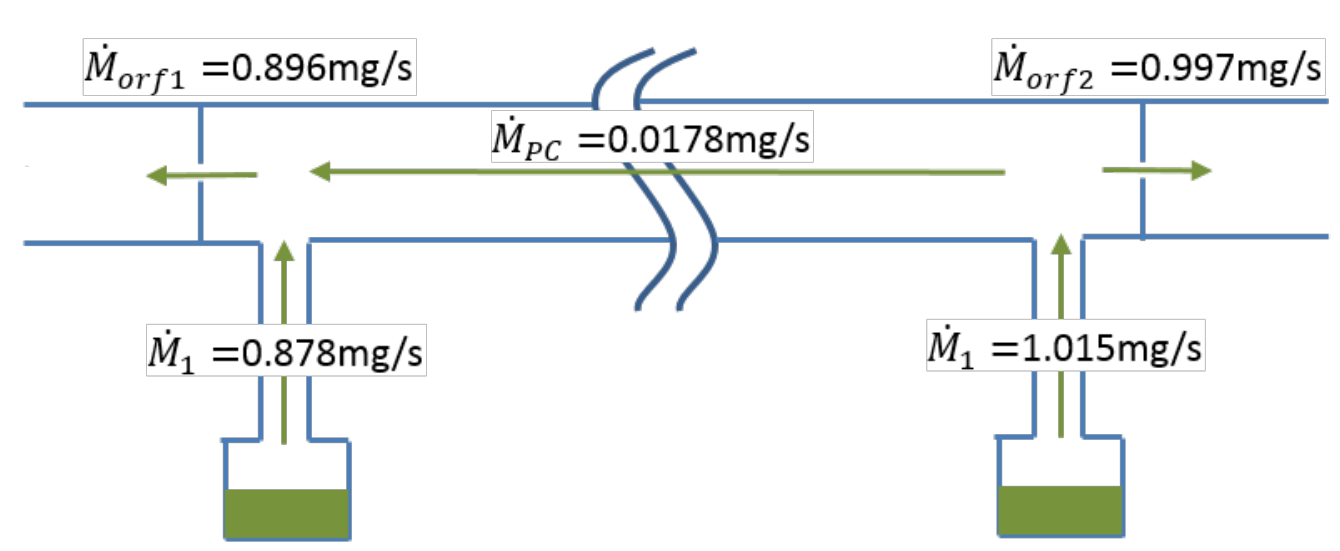}
\caption{Rubidium mass flow rates with a +10\% density gradient along the vapor source (left to right).}
\label{fig:flowsum}
\end{figure}

The first Rb reservoir, situated say at the entrance of vapor source, must provide the Rb mass flow rate to partially compensate for the leak through first orifice minus the flow inside vapor source. %
The flow maintains a +10\% positive density gradient, thus the mass flow rate from reservoir 1 is $\dot M_1$=$\dot M_{orf1}$-$\dot M_{PC}$=0.878\,mg/s ($\dot M_{orf1}$=0.896\,mg/s). %
The second Rb reservoir provides the flow that is equal to the second orifice flow plus the flow inside vapor source, thus the mass flow rate from reservoir 2 is $\dot M_2$=$\dot M_{orf2}$+$\dot M_{PC}$=1.015\,mg/s ($\dot M_{orf2}$=0.997\,mg/s). %
The summary of the mass flow rates is shown on Fig.~\ref{fig:flowsum}.
\begin{figure}[h!]
\centering
\includegraphics[width=0.7\textwidth]{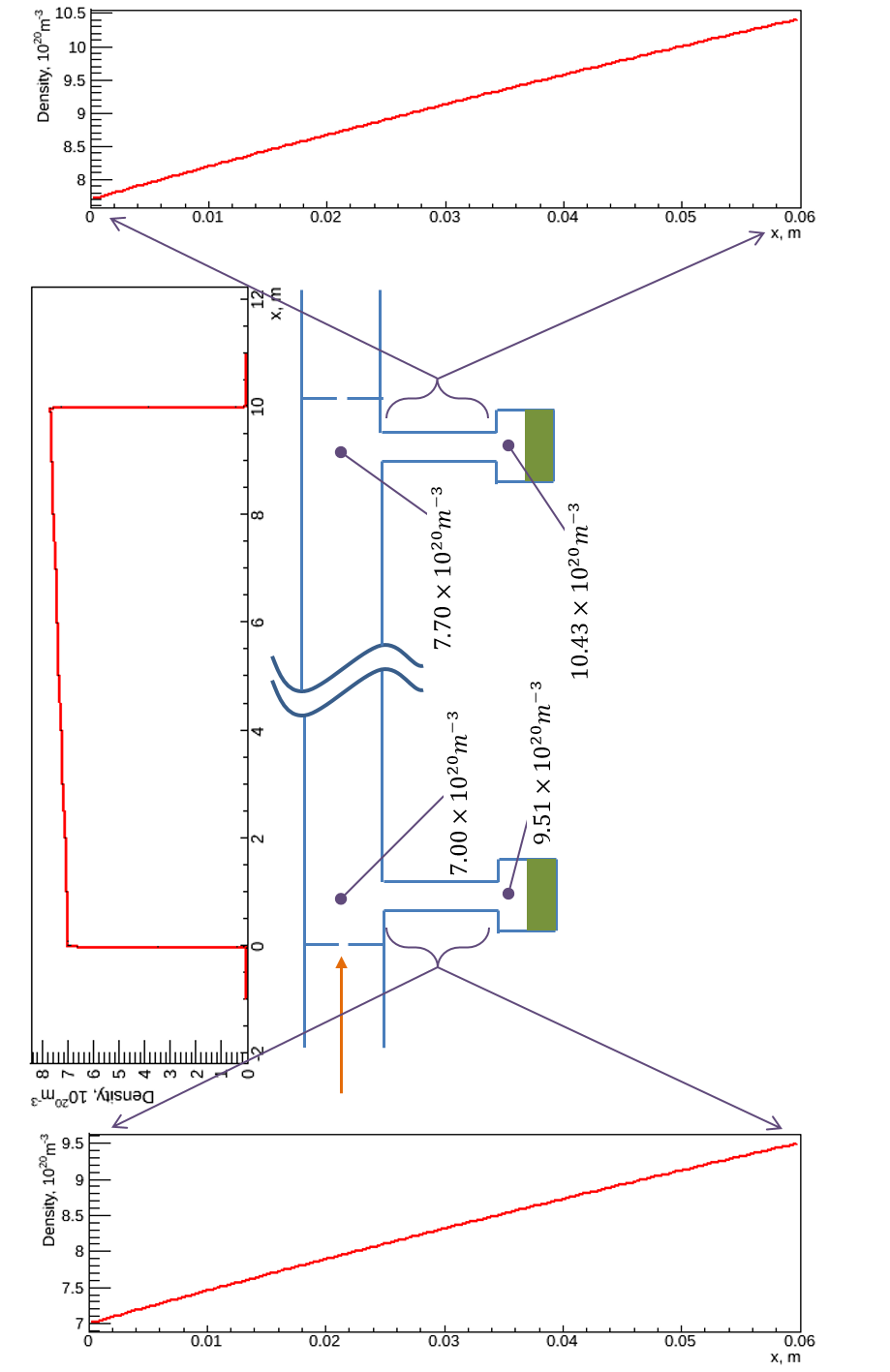}
\caption{Rubidium density distributions for the case of a +10\% density gradient along the source.}
\label{fig:nsumm}
\end{figure}

\subsection{Vapor Density in the Reservoirs}

Knowing the mass flow rate from each Rb reservoir and the density near each reservoir tube, we can calculate the density inside each reservoir. %
Again using Poiseuille law for the rarefied flow regime in the tube between the reservoir and the source:
\begin{equation}
n+n_a= \sqrt{\left( n_{0,1}+n_a \right)^2-\frac{16 \mu \dot M_{1,2} }{\pi a^4\,mk_BT} x}
\end{equation}
As above we assume that the tube connecting each reservoir to the vapor source has length of x=6\,cm and radius of a=1\,cm. %
Therefore, in order to drive the vapor flows, the density in the first and second reservoir must be n=9.51$\times$10$^{14}$ and n=10.43 $\times$10$^{14}$\,cm$^{-3}$, respectively. %
Note that these values are compatible with the simulated ones. %
Figure~\ref{fig:nsumm} shows a summary of the densities in the various part of the vapor source. %

\subsection{Rubidium Evaporation}
The next step is to check whether the required Rb mass flow can be obtained with a reasonable area for the evaporation surface. %
The evaporation rate is given by~\cite{pound72}:
\begin{equation}
J=\frac {\alpha \left( p-p_{sat} \right)}{\sqrt{2 \pi m k_B T}}
\label{eq:evaprate}
\end{equation}
where $\alpha$ is the evaporation (condensation) coefficient. %
For most of the liquids and solids $\alpha$=1.0 with great accuracy. %
The value of $\alpha $ is smaller than one only in the presence of impurities or chemical reactions near the surface of the liquid. %
Here we assume $\alpha$=1. %
The Rb saturation pressure p$_{sat}$ is defined as~\cite{rub85}:
\begin{equation}
log_{10}\left(\frac {p_{sat}[Pa]}{101325} \right)=4.312-\frac{4040}{T[K]}
\end{equation}
One can use these equations to determine the temperature and surface evaporation for each reservoir. %
The advantage of a small evaporation area is a relatively high temperature difference between two reservoirs for the case of a density gradient. %
This could make it easier to control the gradient value. %
The advantage of a large evaporation area is a smaller absolute temperature for each reservoir. %
For example, for an evaporation area of 10\,cm$^2$ the temperature of the reservoirs must be 475.8\,K and 478.3\,K, in order to provide a +10\% gradient at 7$\times$10$^{14}$\,cm$^{-3}$. %

\subsection{Parametric Study}
The main uncertainty in these analytical estimates is the Rb atom diameter and, as consequence, the Rb vapor viscosity. %
To evaluate the sensitivity of the results to this parameter, let us consider two cases: a high viscosity case with a Rb atom diameter of d=396\,pm (100\,pm lower than in the nominal case); and a low viscosity case with d=596\,pm (100\,pm higher than in the nominal case). %

With a Rb atom diameter of d=396\,pm, the viscosity is $\mu$=3.58$\times$10$^{-5}$\,Pa$\cdot$s. %
As a result, the mass flow rates are slightly lower than in nominal case, for the vapor source by around 30\% and by less than 6\% elsewhere. %
The density near the reservoirs is slightly higher, by around 6\%. %
The temperature of both Rb reservoirs is very similar to that in the nominal case.

With a Rb atom diameter of d=596\,pm, the viscosity is $\mu$=1.58$\times$10$^{-5}$\,Pa$\cdot$s. %
The mass flow rates are slightly higher than in nominal case, for vapor reservoir by around 35\% and by less than 5\% elsewhere. %
The density near the reservoirs is slightly lower, by less than 5\%. %
The temperature of both Rb reservoirs is also very similar to that in the nominal case. %

The temperature difference between the Rb reservoirs is similar for all three cases with viscosities varying by a factor of two. %
We conclude that the density gradient in the vapor source can be effectively controlled with a temperature difference between the two Rb reservoirs and is rather insensitive to the Rb vapor viscosity. %

\subsection{Density Gradient Control}
For a fixed geometry, the Rb vapor source has essentially only two available control parameters: the temperatures of each Rb reservoir. %
The temperature of the other parts is determined by these two values. %
For example, the temperature of the source tubes and vapor source should be high enough in order to prevent Rb condensation, which is determined by saturation pressure. %
The temperature in the expansion volume must be low enough to condensate most of the Rb. %

\begin{figure}[h!]
\centering
\includegraphics[width=1.0\textwidth]{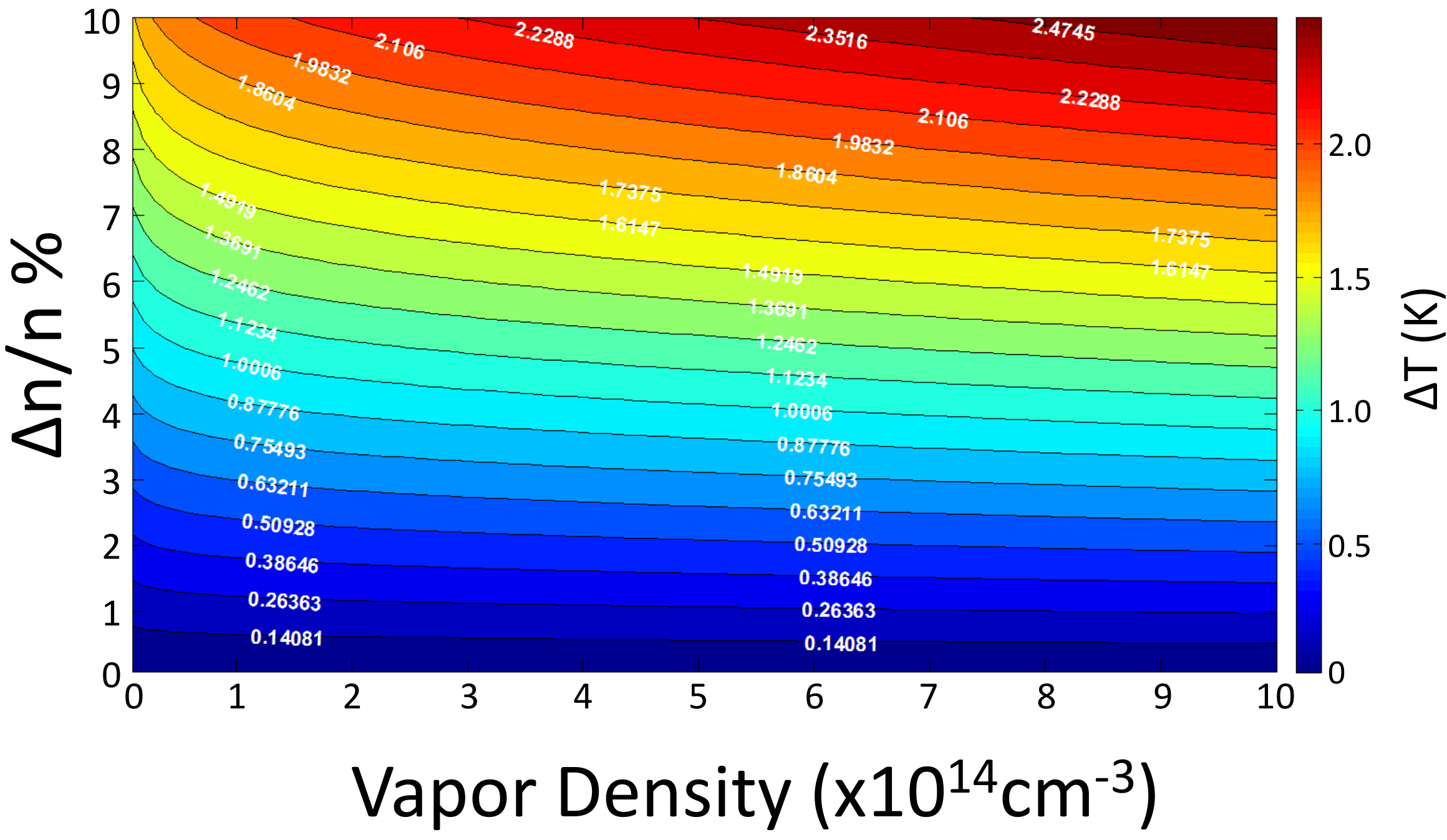}
\caption{The temperature difference between the two Rb reservoirs as a function of density and density gradient in the Rb vapor source (for the reservoir evaporation areas of 10\,cm$^2$). %
The contours are equi-temepratures labelled with the temperature difference.}
\label{fig:deltatsource10cm2}
\end{figure}

The absolute temperature of both reservoirs determines the pressure in the vapor source. %
And the temperature difference between the two reservoirs controls the vapor density gradient along the source tube. 
We therefore plot on Fig.~\ref{fig:deltatsource10cm2} the temperature difference between the two reservoirs to obtain a given density gradient along the source. %
This figure shows that, with an evaporation area of 10\,cm$^2$, one must control the temperature difference with a 0.15\,K precision in order to be able to control the density gradient with a 1\% accuracy, . %

\section{Rubidium Density in the Expansion Volume}

\begin{figure}[h!]
\centering
\includegraphics[width=0.8\textwidth]{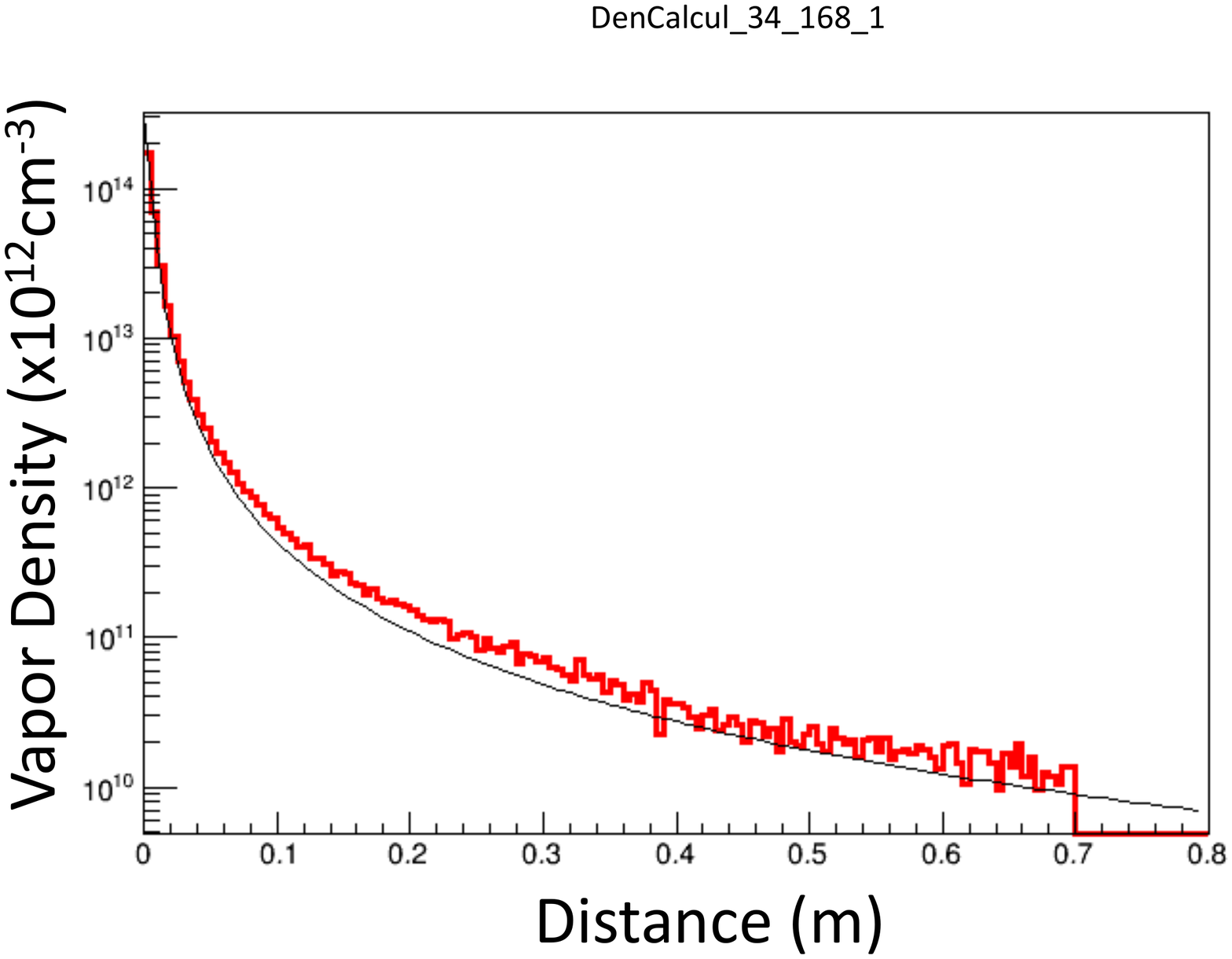}
\caption{Thin black line: plot of Eq.~\ref{eq:densorif}. %
Thick red line: result for the corresponding DSMC with the expansion volume walls are at a temperature of 27$^\circ$C.
}
\label{fig:den1dexpvol}
\end{figure}

Plotting the data of Fig.~\ref{fig:den1ddsmc} with a logarithmic scale (Fig.~\ref{fig:den1dexpvol}) one can see the density profile features in expansion volume. %
As a reminder, the expression for the density is for the expansion of a vapor from an infinite volume at a given density, through an orifice and into an infinite vacuum volume. %
In reality, neither the source nor the vacuum volumes are infinite. %
We therefore use DSMC to determine the density ramp in the AWAKE vapor source case with finite volumes. %
Simulations were performed for perfectly absorbing walls (i.e., Rb atoms stick to the walls with probability one), which physically corresponds to the walls at zero temperature 0\,K. %
\begin{figure}[h!]
\centering
\includegraphics[width=1.0\textwidth]{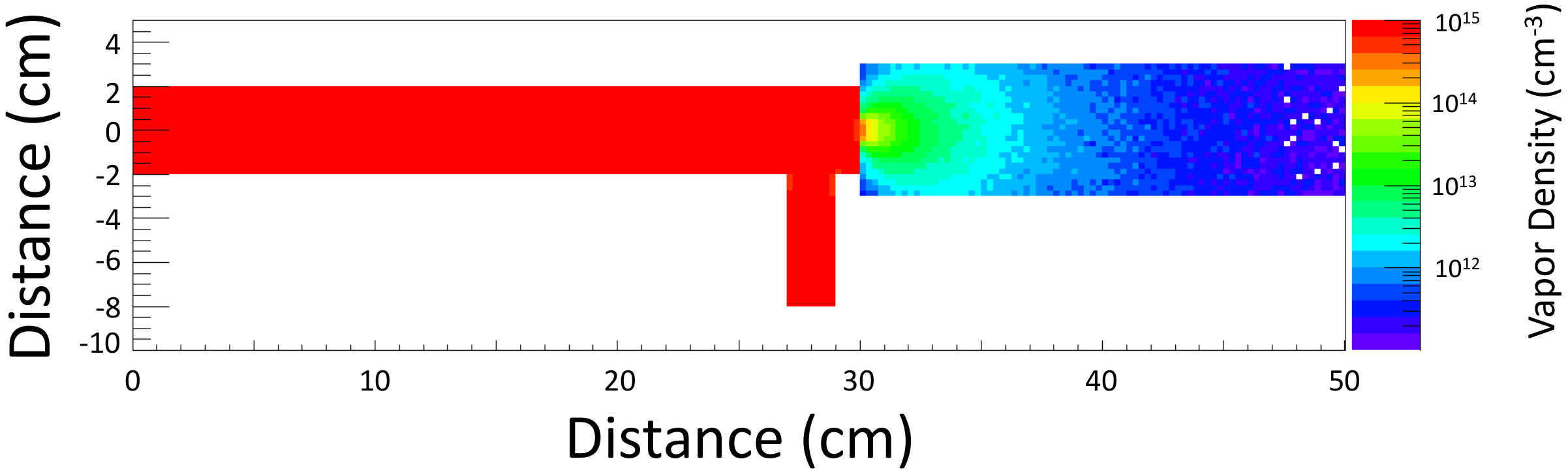}
\caption{DSMC simulation result for the Rb density in the expansion volume. %
Full condensation on all the walls is assumed, except the orifice wall that is kept at a temperature of 500\,K.%
}
\label{fig:den2dexpvol1}
\end{figure}

We plot on Fig.~\ref{fig:den2dexpvol1} the vapor density in the source and in the expansion volume. %
In the expansion volume, the color zones of constant density have an approximately spherical shape tangential to the orifice. %
This indicates that the Rb is emitted from the orifice with a cosine angular law. %
This is in agreement with expectations for the case of an orifice opening small when compared to the expansion volume dimensions. %
Also, as we assume that the walls temperature is 0\,K, the walls of the finite size volume (6\,cm diameter and 20\,cm length in this case) play no role in the density distribution. %

We now formulate the guidelines for a finite size expansion volume acting as a perfect expansion volume. %
First, the walls of the expansion volume must be cold enough to condensate most of the Rb vapor. %
Second, the transition from hot (orifice) to cold (expansion volume walls) temperatures must be on the wall coinciding with the orifice, because this wall is parallel to the Rb flow, where, according to the cosine law, the flow has a minimum (zero). %
The expansion volume must be long enough in other to capture most of the Rb and minimize the losses through the beam pipe that is situated on the wall opposite to the orifice. %
At the same time, the lateral size and shape are not critical, as long as the wall temperatures are cold enough. %
\subsection{Expansion Volume Walls Temperature}
From Fig.~\ref{fig:den1dexpvol} (black line), we read that 0.5\,m away from the orifice (and it the case of flow between infinite volume), the on-axis density is 1.75$\times$10$^{10}$\,cm$^{-3}$. %
This density corresponds to the saturation pressure of the Rb vapor at 28$^\circ$C. %
Thus, by keeping the temperature of the expansion volume walls lower than this value, the evaporation from the walls will contribute a negligible additional density when compared to the perfectly absorbing wall case. %

To verify this result, a DSMC simulation of the expansion volume was performed for the case of a wall temperature at 27$^\circ$C. 
The Rb flow from the walls is simulated according to Eq.~(\ref{eq:evaprate}). %
We plot on Fig.~\ref{fig:den1dexpvol} the on-axis density obtained from the simulation. %
For this wall temperature the Rb density ramp obtained from the simulation is in good agreement with the analytical result that neglects the evaporation from walls (perfectly absorbing walls). %
Therefore, a finite volume with a finite temperature, below the condensation temperature if Rb, leads to a density gradient very similar to the ideal case. %

\section{Density Step in the Vapor Source}

\begin{figure}[h!]
\centering
\includegraphics[width=0.8\textwidth]{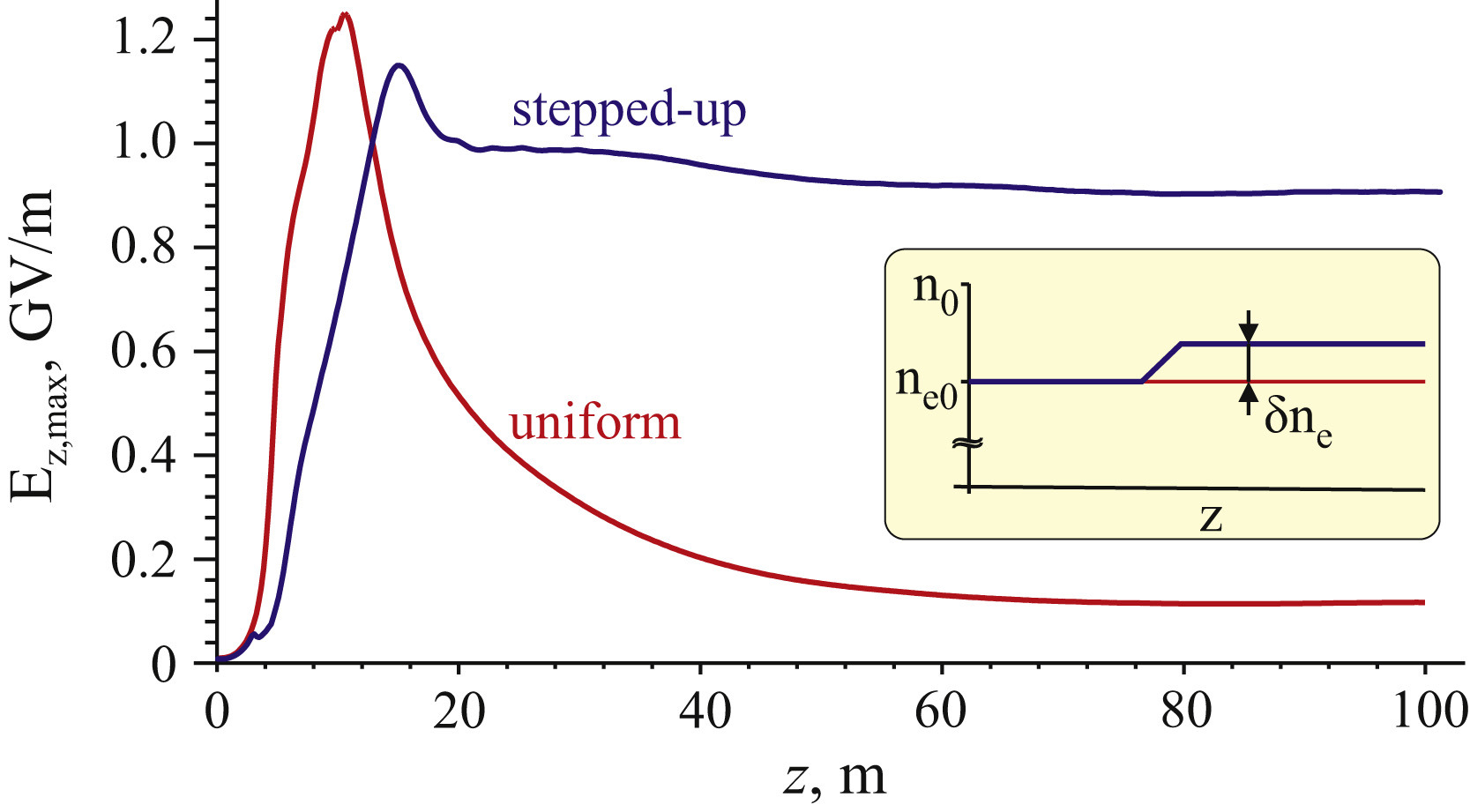}
\caption{The maximum wakefield amplitude versus the propagation distance for the stepped-up and uniform plasmas for a simulation with an LHC bunch. The inset illustrates the changes in the plasma density profile at z=3\,m. %
Figure is taken from~\cite{Caldwell16}.}
\label{fig:LotovDenStep}
\end{figure}

It was shown in reference~\cite{Caldwell16} that a small density step, on the order of 3\% over a few centimeters along the vapor column, can significantly improve the average acceleration gradient for the proton driven plasma wakefield acceleration (Figure \ref{fig:LotovDenStep}).

Here we consider two ways to produce a small density step: first, by using an internal orifice in the presence of flow inside the vapor source, i.e., with a density gradient, second, by using a wall temperature step along the source. %
In order to characterize the density step produced by these two methods, DSMC simulation were performed. %
In all simulations presented in this section, we used the baseline density of 7$\times$10$^{14}$\,cm$^{-3}$. %

\subsection{Density Step from an Aperture}

In the presence of flow along the vapor source, we can create a density step with an orifice placed inside the source tube. %
Because of the orifice, more particles will accumulate on one side of the orifice, on the side the flow is coming from. %
In steady state, the density gradients before and after orifice, and due to the flow, will be equal: the mass flow is the same and the density step is much lower than the absolute density. %
There is no simple analytical expression for the density profile in this case. %
Also, the step value depends on the mass flow or equivalently on the density gradient along the vapor source. %
The technical implementation is very simple, the density step length can be relatively small, again on the order of orifice diameter, as shown above. %

\begin{figure}[h!]
\centering
\includegraphics[width=0.8\textwidth]{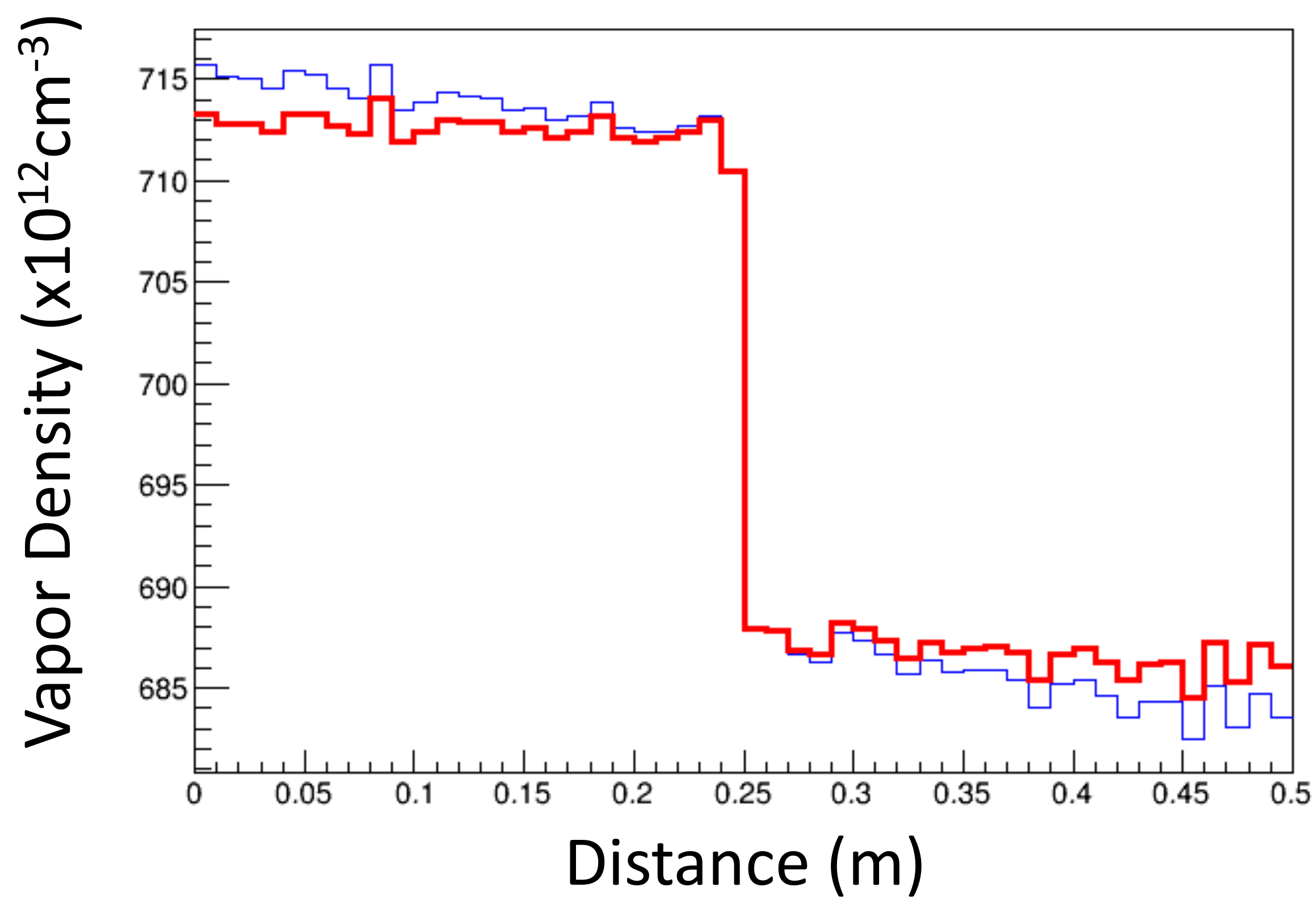}
\caption{Simulation result for the density step created by a 1\,cm diameter orifice, placed at the 0.25\,m location, for the baseline vapor density with the density gradient of +10\%. %
The blue curve represents the simulated density profile and the red curve the density profile from which the global density gradients were subtracted in order to better see the step. %
The calculated density step is around 3.7\% around 7$\times$10$^{14}$\,cm$^{-3}$.}
\label{fig:Den1cm}
\end{figure}

Assuming that from physical limitations, such as the clearance needed for the proton transverse beam size, the orifice diameter cannot be smaller than 1\,cm, we determine the highest possible density step that can be reached with the large density gradient of +10\%. %
Here we simulate only an infinitely thin orifice. %
We can create a larger density step with a thicker orifice, but the density step extent would also be longer. %
The simulation results are presented in Fig.~\ref{fig:Den1cm} and show that a density step of 3.7\% can be achieved. %

\subsection{Density Step from a Temperature Step}
\begin{figure}[h!]
\centering
\includegraphics[width=0.8\textwidth]{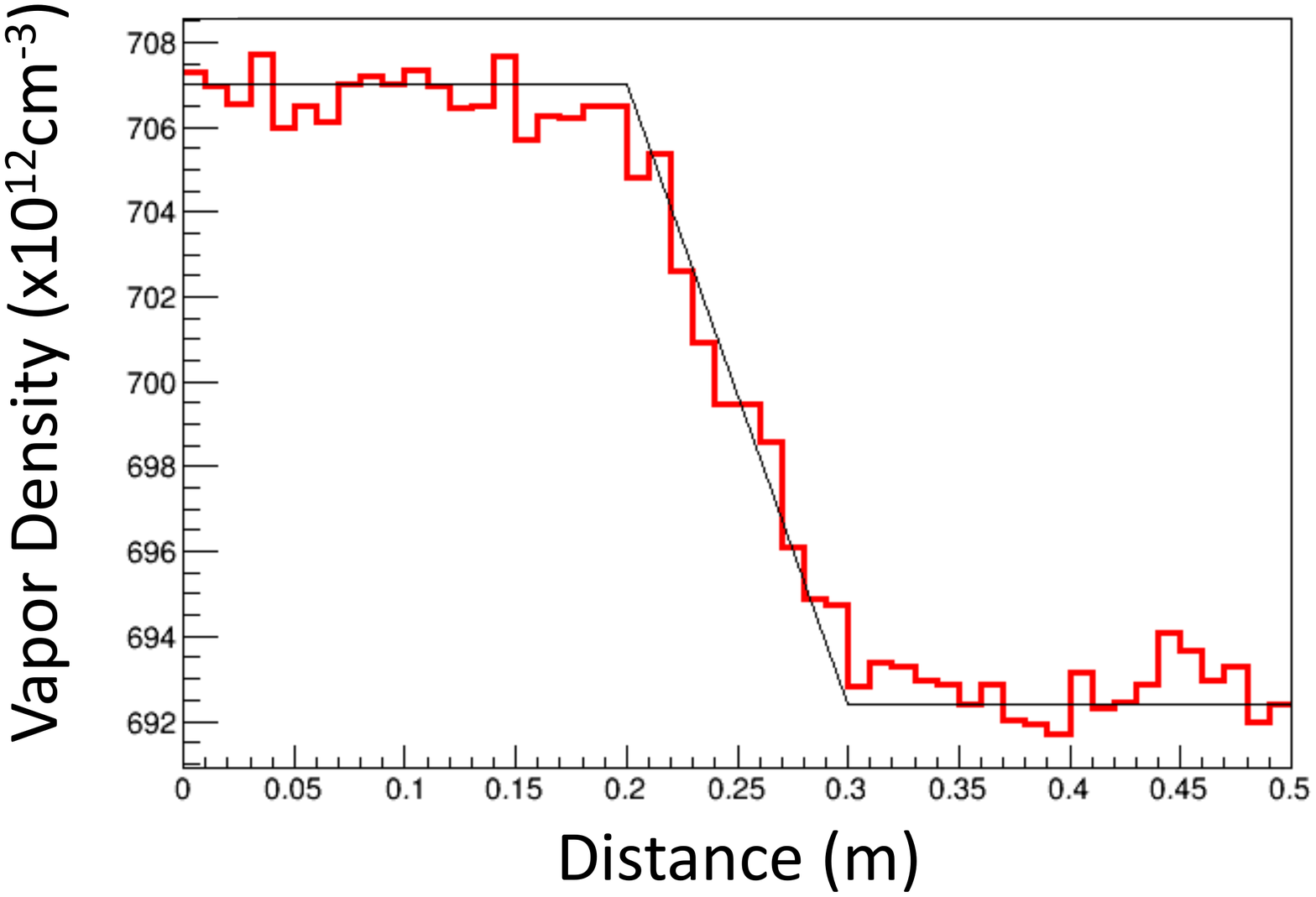}
\caption{Simulation result for the density step created by a 10$^\circ$C temperature difference along the source for the baseline density with no density gradient (no flow). %
The temperature linearly increases over 10\,cm, from 473\,K (at $<$0.2\,m) to 483\,K (at $>$0.3\,m). %
The black line shows the density calculated from the state equation for the local temperature of the walls. %
The calculated density step is around 2.1\%.}
\label{fig:Den10degC}
\end{figure}

When two parts of the vapor source have slightly different temperatures, the gas density is mofified according to the state equation $p=nk_BT$, that is the local density is inversely proportional to the local temperature. %
This configuration allows for a simple means to control the density step. %
The step value does not need and is not affected by a flow along the vapor source. %
However, the the length of the density step can be quite long, on the order of the extent of the wall temperature variation. %
Figure~\ref{fig:Den10degC} shows a density step of 2.1\% created with a temperature step of 2.1\%, from 473\,K to 483\,K.

Another interesting point is to determine how the temperature driven density step is affected by the flow inside vapor source. %
We used the same conditions as in the previous simulation and added the vapor source flow that corresponds to a +10\% density gradient, with a mass flow of 0.0178\,mg/s along the source and the value of density step remains unchanged. %

\section{Conclusions}
We have presented the scheme for a Rb vapor source that is used as a plasma source in the AWAKE plasma wakefield acceleration experiment. %
The plasma wakefield acceleration process imposes a number of stringent parameters on the plasma: %
electron density adjustable in the (1-10)$\times$10$^{14}$\,cm$^{-3}$ range; %
a 0.25\% relative density uniformity; %
sharp, $<$10\,cm, density ramps at each end; %
a density gradient adjustable from -3 to +10\% over 10\,m; %
and a \%-level density step near the beginning of the plasma column. %
We showed with analytical and direct simulation Monte Carlo results that the Rb density in the proposed source meets these requirements. %
Additional calculations with the finite-elements code COMSOL and the Test-Particle MC code Molflow+ have also been used to validate the results obtained. %
Laser ionization then directly transfers the neutral vapor parameters reached by the vapor to the electron plasma density. %
Therefore the source satisfies the requirements for the plasma. %
Self-modulation and acceleration experiments will be the ultimate test of the plasma source parameters. %



\newpage

\begin{thebibliography}{99}

\bibitem{bib:chen}Chen P. , et al., 1985 {\it Phys. Rev. Lett.} {\bf 54} 693 

\bibitem{bib:caldwellshort}Caldwell A. et al., 2009 {\it Nature Physics} {\bf 5}, 363 

\bibitem{bib:kumar}Kumar N. et al., 2010 {\it Phys. Rev. Lett.} {\bf 104}, 255003 

\bibitem{bib:ozdensity}{\"O}z E., Muggli P., A novel Rb vapor plasma source for plasma wakefield accelerators, \emph{Nucl. Instr. Meth.} A {\bf 740}, 197 (2014), E. {\"O}z et al., {\it Nucl. Instr. and Meth. in Phys. Res. A} {\bf 829}, 321 (2016)

\bibitem{bib:AWAKE}AWAKE Collaboration, 2014 {\it Plasma Phys. Control. Fusion} {\bf 56}, 084013; 
Gschwendtner E. 2016, {\it Nucl. Instr. and Meth. in Phys. Res. A} {\bf 829}, 76; 
Caldwell A. \etal, 2016 {\it Nucl. Instr. and Meth. in Phys. Res. A} {\bf 829}, 3; 
Muggli P. \etal, 2017 submitted to {\it Plasma Phys. and Contr. Fusion} 
\url{https://arxiv.org/abs/1708.01087}


\bibitem{Caldwell16}Caldwell A. , et al., 2016 \emph{Nuclear Instruments and Methods in Physics Research Section A} {\bf 829} 35

\bibitem{Lotov13}Lotov K.V., Pukhov A., and Caldwell A., 2013 Effect of plasma inhomogeneity on plasma wakefield acceleration driven by long bunches, \emph{Phys. Plasmas} {\bf 20}, 013102 

\bibitem{bib:vieiraion}Vieira J. , et al., 2012 {\it Phys. Rev. Lett.} {\bf 109}, 145005 

\bibitem{bib:dsmc}Bird G. A., Molecular Gas Dynamics and the Direct Simulation of Gas Flows, Claredon, Oxford (1994)

\bibitem{kersevan} Kersevan R., Pons J.-L., 2009 ÒIntroduction to Molflow+Ó, Journal of Vacuum Science and Technology A 27, 1017 
; \url{https://molflow.web.cern.ch/} 

\bibitem{kersevan2} Kersevan R., presentation at the AWAKE Physics Board Meeting, 13 Feb. 2015

\bibitem{danilatos}Danilatos G., 2000\emph{Rarefied Gas Dynamics. 22nd International Symposium, Sydney (ed. T.J.Bartel and M.A.Gallis), AIP Conference Proceedings}, {\bf 585}, 924 

\bibitem{sharipov01}Sharipov F., \emph{Rarefied Gas Dynamics, edited by T. J. Bartel and M. A. Gallis, 22nd
Int. Symp., Sydney}, 494 (2001)

\bibitem{sharipov98}Sharipov F., Seleznev V., 1998 \emph{J. Phys. Chem. Ref. Data} {\bf 27}, 657 

\bibitem{sharipovCERN}Sharipov F., \emph{CAS - CERN Accelerator School and ALBA Synchrotron Light Facility: Course on Vacuum in Accelerators}, \url{dx.doi.org/10.5170/CERN-2007-003.1} (2006)

\bibitem{pound72}Pound G., 1972 \emph{J. Phys. Chem. Ref. Data} {\bf 1}, 135 

\bibitem{rub85}Steck D., 2013 Rubidium 85 D line data, steck.us/alkalidata/rubidium85numbers.pdf 



\end{thebibliography}
\end{document}